# Inventions on keyboard illumination
## -A TRIZ based analysis

**Umakant Mishra**

Bangalore, India

umakant@trizsite.tk

http://umakant.trizsite.tk

## Contents

## 1. Introduction

The conventional computers are not usable in dark as the user cannot see the keyboard to operate properly. But there are many situations where the user may like to work on his computer under low light conditions, such as, during nights, while traveling in flights/ trains having low lights, working in the garden during evening etc. So it is necessary to provide some light to the laptop to operate the keyboard under low light situations.

There are several methods of providing light to the keyboard. Some of the keyboards use light diffusers that provide lighting across the area of the keyboard. Other types use electro-luminescent lighting components. Some common problems in illuminating a keyboard are as follows.
- Some solutions using electro-luminescent materials increase the cost of the keyboard.
- Some solutions use more battery.
- Some solutions (the light being inside the keyboard) generate more heat.
- Some other solutions are complex in structure.
- Some light sources don't provide uniform lighting.

There is a need for a better illumination system that can avoid the problems mentioned above.

## 2. Inventions on keyboard illumination

### 2.1 Computer keyboard light system (Patent 5868487)

**Background problem**

With the increasing use of computers, more number of PCs are found outside the office such as in dormitories and bathrooms etc. where lighting is found to be a problem. It is necessary to keep a lighting system for convenience in using the keyboard.

**Solution provided by the invention**

Polley, et al. invented a lighting system for the keyboard (Patent 5868487, assigned to Catalina Lighting, Feb 99), which uses a small lamp mounted on a flat plate on the computer keyboard or pc monitor. The lamp arms are telescopeable and are rotatable by rachet connections to allow for adjustment of the lights. The arms could also be adjusted to illuminate either the keyboard, or monitor screen or a copyholder as required.

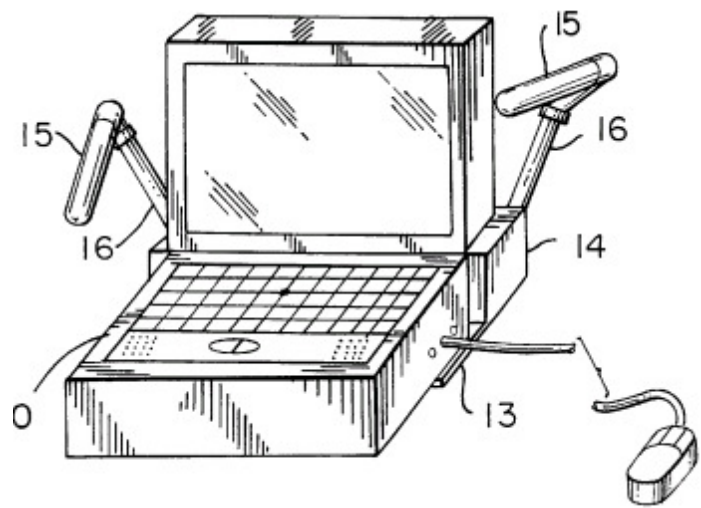

**TRIZ based analysis**

The keyboard should illuminate itself **(Ideal Final Result)**.

The invention provides a lighting system with the keyboard **(Principle-8: Counterweight)**.

The lamp arms are flexible to change the lighting position **(Principle-15: Dynamize)**.

**2.2 Keyboard illuminator and a method for illuminating a keyboard (Patent 6145992)**

**Background problem**

When there is not sufficient amount of light available, a user is unable to see the keyboard and unable to accurately and efficiently operate the computer. The conventional computers and portable laptops are of no use when the user cannot see the keyboard. There is a need for the user to have full visibility of the keyboard.

**Solution provided by the invention**

Wattenburg disclosed a keyboard illuminator (Patent 6145992, assignee- Wattenburg, Nov 2000), which illuminates the computer keyboard by utilizing the light emitted from the display screen of the computer.

The keyboard illuminator includes a rigid planar member that is at least partially transmitting and at least partially reflecting and an apparatus for mounting the member on the housing. The member is mounted so that the member is disposed at an angle to the display screen between the display screen and the user such that at least a portion of light emitted by the display screen is reflected onto the keyboard to illuminate the keyboard for operation.

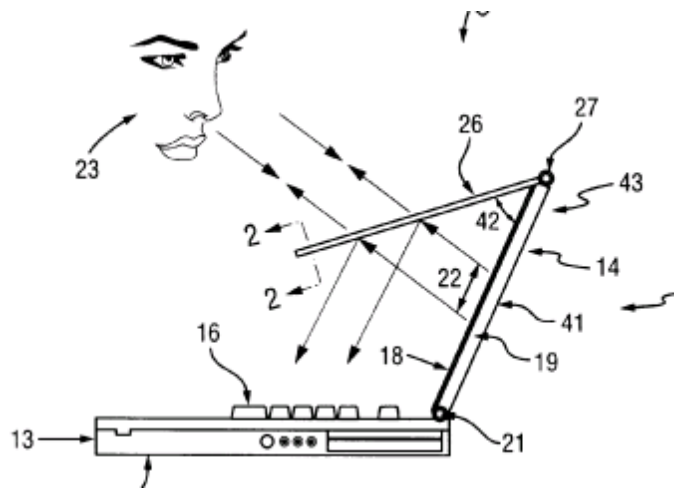

**TRIZ based analysis**

Display screens are designed to emit sufficient light when viewed in well-lit locations. Thus when viewing the display screen in low light conditions, the display screen emits much more light than it is required for effective viewing. The invention reflects a portion of this excessive light to illuminate the keyboard. **(Principle-22: Blessings in disguise)**.

The invention uses the already available light without using external light sources and without consuming additional power **(Principle-25: Self service)**.
The invention uses a partially reflecting layer which is coupled to the side of the display screen and is adjustable to any angle that is suitable for the user **(Principle-15: Dynamize)**.

**2.3 Backlighting for computer keyboard (Patent 6322229)**

**Background problem**

It is difficult to operate on the keyboard under low light conditions. Laptop operators often need lighted keyboards while working inside aeroplane or low light environments. There is a need for a effective and economical backlighting system for the keyboards. Besides it's comfortable to have a lighted keyboard that matches with the luminescence of the monitor.

There are a few inventions on illuminating a keyboard. One method keeps the light below the keyboard and allows light to come through the key stems. The keys are clear, transparent or translucent to allow lights to pass through. This method limits the light path, consumes more power and generates more heat. There is a need for an effective backlighting system.

**Solution provided by the invention**

Chan et al. invented a backlighting system (Patent 6322229, assigned to Questech International, Nov 01) for the keyboards used with laptops, desktops, Internet-TVs etc. The light-emitting panels are exterior to the key-switches of the keyboard. The panels are made of electro-luminescent material. The EL panels are powered by the computer keyboard port, via voltage inverters. The key switch is preferably translucent. The light from the light-emitting panels radiates upward through, and around the translucent keycaps.

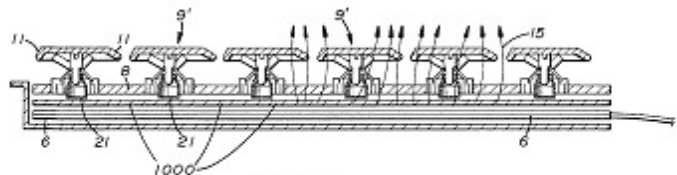
FIG. 5

**TRIZ based analysis**

Lighting from outside the keyboard requires external lamps and causes inconvenience in carrying those lighting accessories. Lighting from inside the keyboard blocks most part of light while coming through the keyboard and heats the system **(Contradiction)**.

The invention uses lighting from outside the keyboard frame but under the keys. The panels are made of electro-luminescent material. The key switch is also translucent **(Principle-32: Color change)**.

**2.4 Thin light permeable keyboard multiple switch assembly including scissors type actuator mechanisms (Patent 6545232)**

**Background problem**

There is a need to illuminate the keyboard to operate in dark or less light situation. There are different mechanisms to illuminate the keyboard, such as lighting from behind, lighting from the top, using LEDs and so on. Some of them use more battery, some of them generate more heat, and some others are complex in structure. There is a need to illuminate the keyboard in an easier way.

**Solution provided by the invention**

Huo-Lu invented a keyboard (patent 6545232, assigned to Sunrex Technology, April 2003) which is thin and light permeable. The keyboard uses a luminescence board, a film circuit board and a base board. The keys are also light permeable. The characters on the keys are light impermeable to make the characters visible.

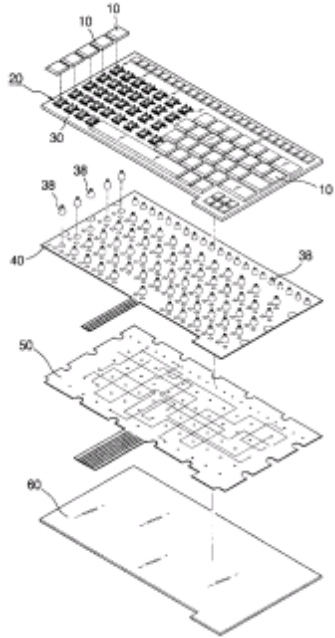

The keyboard contains a frame board, a luminescence board, a film circuit board, and a base board, all of which are light permeable. This facilitates the user operating the computer in dark site.

## TRIZ based analysis

The invention uses thin and light permeable frame board and circuit boards **(Principle-30: Thin and flexible)**.

Light permeable frame board and key switches are used for luminescence **(Principle-32: Color change)**. The keys are light permeable and the characters on the keys are light impermeable to make the characters visible.

### 2.5 Backlit keyboard (Patent 6590508)

**Background problem**

It is necessary to provide some light to operate the keyboard under low light situations. Some of the keyboards use light diffusers that provide lighting across the area of the keyboard. Other types use electro-luminescent lighting components. But electro-luminescent materials are expensive which increase the cost of the keyboard. There is a need for a backlit keyboard that provides uniform illumination and ease of manufacturing.

**Solution provided by the invention**

Howell Bryan invented a backlit keyboard (Patent Number: 6590508, July 2003), which solves most of the problems in the prior art. According to the invention, the keyboard and illumination subassemblies are independent and separately manufactured. The light panel is mounted on the keyboard subassembly. Each key is mounted with a key cap. The key cap includes a body formed of a non-opaque material having an opaque character indicia portion forced thereon. Light is transmitted through body such that the character indicia portion may be viewed in low light conditions. The key cap may include a light diffuser to enhance the projection of light from the light panel to the character indicia portion. Each key will extend through a perspective opening in the light panel. With this method, the entire area of the keyboard is illuminated more uniformly than other methods.

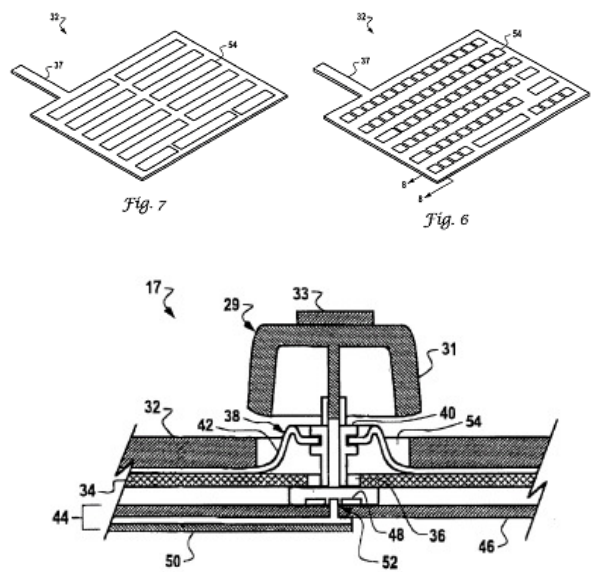

**TRIZ based analysis**

The invention makes the keyboard assembly and illumination subassembly independent components to be separately manufactured **(Principle-1: Segmentation)** for ease of manufacturing.

### 2.6 Computer with keyboard illuminator (Patent 6561668)

**Background problem**

The laptop keyboard needs some light to be operated in dark and low light conditions. There are several inventions earlier to build illuminated keyboards using electro-luminescent materials. Some inventions used detachable light source. Although some patents suggest placing the light at different places on a laptop, no specific position is really convenient for the purpose.

**Solution provided by the invention**

Katayama et al. disclosed a proper position of placing the light source (Patent 6561668, Assigned to IBM, May 03) to find a proper position of placing the light source. According to the invention the LED holder is provided in the upper portion of an LCD and an LED is attached inside the LED holder. Light emitted from the LED passes through an aperture provided in the bottom portion of the LED holder and illuminates the keyboard. Furthermore, switching on or off the LED is manually performed by a switch installed in the portable computer and is also controlled from a utility program.

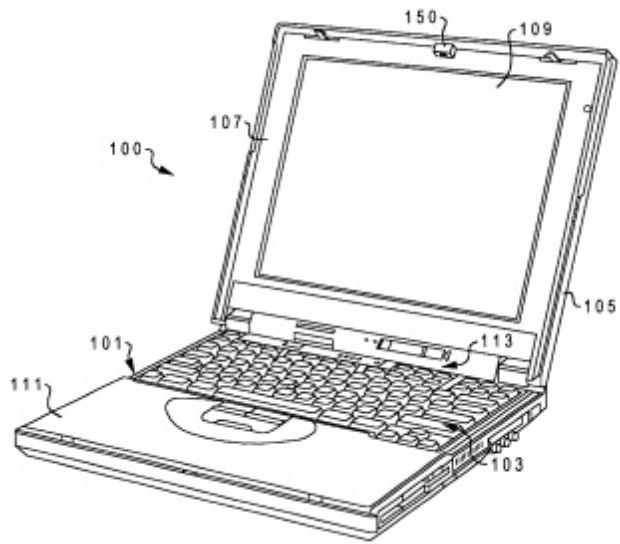

By placing the light source on the top, the entire keyboard is effectively illuminated. The LED holder has a shutter for preventing light from extending beyond the main body and illuminating unwanted ranted. The user can adjust the degree of opening of the aperture as required to focus the light on the keyboard.

**TRIZ based analysis**

The invention keeps the light source above the keyboard instead of keeping behind the keyboard **(Principle-17: Another dimension)**.

The light can be switched on or off manually when required **(Principle-15: Dynamize)**.

## 3. Findings and conclusion

Luminescence of the keyboard is a desirable feature for portable computers. There may be several methods to illuminate the keyboard, the most crude may be carrying an external light and fixing at a point where it can light the keyboard. But solutions like this are not very convenient. The inventors have been looking for better solutions.

A good illumination system do not consume more battery, do not increase the size and weight of the laptop, do not increase heat inside the laptop box, and do not cost much to manufacture. The light should uniformly illuminate the whole keyboard and should not throw light on other areas that are not required.